\documentclass[aps,pre,preprint,showkeys]{revtex4-1}
\pdfoutput=1
\usepackage{graphicx}
\usepackage{amsmath,amsthm}
\usepackage{verbatim}
\usepackage{enumitem}

\newcommand*{\citen}[1]{%
  \begingroup
    \romannumeral-`\x 
    \setcitestyle{numbers}%
    \cite{#1}%
  \endgroup  
}

\begin{document}

\title{Phase Separation and Self-Assembly in a Fluid of Mickey Mouse Particles}

\author{Guido Avvisati}
\affiliation{Debye Institute for Nanomaterials Science, Utrecht University, Princetonplein 1, 3584CC Utrecht, The Netherlands}
\author{Marjolein Dijkstra}
\email[]{m.dijkstra1@uu.nl}
\affiliation{Debye Institute for Nanomaterials Science, Utrecht University, Princetonplein 1, 3584CC Utrecht, The Netherlands}

\date{\today}

\begin{abstract}
  Recent developments in the synthesis of colloidal particles allow for control over shape and inter-particle interaction. One example, among others, is the so-called ``Mickey Mouse'' (MM) particle for which the self-assembly properties have been previously studied yielding a stable cluster phase together with elongated, tube-like structures. Here, we investigate under which conditions a fluid of Mickey Mouse particles can yield phase separation and how the self-assembly behaviour affects the gas-liquid coexistence. We vary the distance between the repulsive and the attractive lobes (bond length), and the interaction range, and follow the evolution of the gas-liquid (GL) coexistence curve. We find that upon increasing the bond length distance the binodal line shifts to lower temperatures, and that the interaction range controls the transition between phase separation and self-assembly of clusters. Upon further reduction of the interaction range and temperature, the clusters assume an increasingly ordered tube-like shape, ultimately matching the one previously reported in literature. These results are of interest when designing particle shape and particle-particle interaction for self-assembly processes.
\end{abstract}


\maketitle

\section{Introduction}
One of the fundamental questions in condensed matter physics is to understand the relation between the microscopic and macroscopic properties of a system. The answer to this question is also closely related to the possibility of creating new routes to fabricate novel functional materials. A variety of fabrication protocols has been devised over the past years, but more recently the focus has been shifted towards encoding information for the self-assembly into the basic colloidal building blocks\cite{bib:agostiano-fabrication.persp,bib:manners-self.perspective,bib:granick-colloidal.assembly}. This paradigm is now known as self-assembly or ``colloidal LEGO''. Beforehand, it is unknown how the encoded information will manifest itself at the end of the self-assembly process, and it is even unknown what kind of encoded microscopic information will yield the desired macroscopic structure. Answering these questions is the key to using the self-assembly process of colloidal particles as a new method to fabricate functional materials.

``Hard'' particle systems, which are purely entropic, give insights into the role of the particle shape in the self-assembly process. It has been shown that entropy alone can give rise to a large variety of close-packed crystal structures, as well as liquid and plastic crystal phases \cite{bib:sacanna-review.colloids,bib:solomon-review.colloids,bib:basavaraj-review.colloids,bib:glotzer-anisotropic.assembly,bib:damasceno-shape.assembly,bib:kuijk-rods.assembly,bib:sacanna-shape.anisotropic,bib:sacanna:review.shape.anisotropy,bib:lee-shape.anisotropy}. On the other hand, having particles explicitly attract each other along specific directions can be of advantage to target particular open structures as, for instance, diamond crystals and kagome lattices\cite{bib:romano-diamond.xtal,bib:chen-kagome,bib:granick-open.lattices}. This has brought a considerable amount of interest to what are now called patchy colloids, particles that interact with each other only via specific spots located at the particle's surface\cite{bib:sacanna-review.patchy,bib:kretzschmar-review.nanomaterials,bib:solomon-spheroidal.patchy,bib:pine-colloidal.valence,bib:bianchi-review.patchy}. Directional interactions can be given to colloids for example via attaching complementary DNA strands to the surface, as in Refs. [\citen{bib:pine-colloidal.valence,bib:gang-dna.assembly}].

Another way of inducing directional interactions between building blocks is to combine depletion forces and surface roughness asymmetry\cite{bib:badaire-shape.selectivity.1,bib:badaire-shape.selectivity.2,bib:kraft-colloidal.molecules,bib:kraft-pacci.anisotropic}. In fact, if a building block is made up of rough and smooth lobes, the addition of depletants to the solution will induce specific attractions between the smooth lobes of different particles, while producing repulsions between the rough-smooth lobes and the rough-rough lobes\cite{bib:kraft-colloidal.molecules,bib:kraft-pacci.anisotropic}. This approach has been used to self-assemble dumbbells\cite{bib:kraft-dumbbells} and, more recently, ``Mickey Mouse'' (MM) particles -- trimers with one big smooth lobe (also referred to as the ``head'' of the MM particle) and two smaller rough lobes (the ``ears'')\cite{bib:avvisati-mickey.mouse}. In particular, it has been shown that the MM particles self-assemble, at interaction strength of $\sim 9-10 k_{\textrm{B}}T$ into tube-like structures\cite{bib:avvisati-mickey.mouse}.

Yet, with respect to isotropically attractive spheres like, for example, Lennard-Jones particles or Hard-Sphere Square-Well (HSSW) particles, that exhibit a gal-liquid (GL) phase separation, it is so far unclear whether MM particles can also undergo an analogous gas-liquid phase separation (colloidal poor-colloidal rich) and, if this is the case, what is the interplay of the latter with the depicted self-assembly scenario. The aim of this work is to address this question with Monte Carlo (MC) simulations. 
To do so, we follow the evolution of the GL binodal line for a series of models connecting the isotropic HSSW spheres to the MM particles. The path proceeds firstly by increasing the rough smooth (ear-head) bond length until we reach the experimental particle geometry, and secondly by decreasing the interaction range until the experimental one is reached, as described in Ref. \citen{bib:avvisati-mickey.mouse}.

The paper is organised as follows: we introduce the model and the path connecting HSSW spheres and MM particles in Sec. \ref{sec:model}, where we also discuss details about the different computational techniques we have used. In Sec. \ref{sec:res}, we present our results on the phase separation of MM particles and the transition to self-assembly. We also show how the particle geometry affects the structure of the liquid. We summarise and discuss these results in Sec. \ref{sec:summary}, where we also provide an outlook for future research directions.

\section{\label{sec:model}Model and Methods}
\begin{figure}[htb]
  \centering
  \includegraphics[scale=2.7]{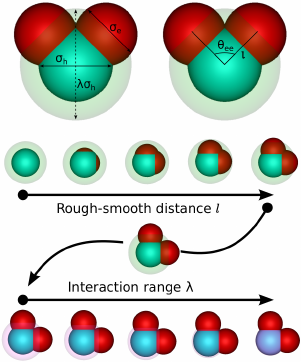}
  \caption{{\small Model and parameters considered in this work. An experimental MM particle is obtained by a SW sphere by first changing the distance $l/\sigma_h$ between the rough and the smooth beads (ear-head distance) and then by decreasing the interaction range $\lambda$}. The red beads (rough protrusions) have diameter $\sigma_e$, while the blue bead (smooth protrusion) has diameter $\sigma_h$.}
  \label{fig:mm_model}
\end{figure}

We begin with the geometry of the MM particle as shown in Fig. \ref{fig:mm_model}. Each MM particle is represented by an aggregate of three spheres, two small spheres (``ears''), of diameter $\sigma_e$, which represent rough lobes interacting with steric repulsions and a third bigger sphere (``head''), of diameter $\sigma_h$, which corresponds to the smooth lobe and plays the role of an attractive site. The interaction $u_{i,j}$ between a pair of MM particles $i$ and $j$ depends on the positional and orientational degrees of freedom $\boldsymbol{r}_i,\boldsymbol{r}_j,\boldsymbol{\Omega}_i,\boldsymbol{\Omega}_j,$ respectively, which we drop here to lighten the notation, and consists of an attractive and a repulsive contribution, 
\begin{equation}
  u_{ij} = u^{\mathrm{att}}_{ij} + u^{\mathrm{rep}}_{ij}
  \label{eq:int.en}
\end{equation}

The attractive part $u^{\mathrm{att}}_{ij}$ acts between the larger beads of particle $i$ and $j$, and is given by a square-well (SW) interaction,
\begin{equation}
  \beta u^{\mathrm{att}}_{ij} = \beta u^\mathrm{SW}(r^{hh}_{ij})=
  \begin{cases}
    \beta\varepsilon & \mathrm{for}\;\; \sigma_{h\phantom{h}} \leq r^{hh}_{ij} \leq \lambda\sigma_h\\[0.3em]
   0 & \mbox{for}\;\; r^{hh}_{ij} > \lambda\sigma_h
  \end{cases} 
  \label{eq:sw.part}
\end{equation}
where $\beta\varepsilon = \varepsilon/k_{\mathrm{B}}T$, $\epsilon < 0$, represents the interaction strength compared to the thermal energy, $\lambda$ is the interaction range, and $r^{hh}_{ij}$ is the center to center distance from the head of MM particle $i$ to the head of MM particle $j$.

The repulsive part $u^{\mathrm{rep}}_{ij}$, is determined by the hard-core interactions between the heads ($h$) and the ears ($e_1,e_2$) of two MM particles, and reads
\begin{equation}
  \beta u^{\mathrm{rep}}_{ij} = \sum_{\alpha,\beta = h,e_1,e_2}\beta u^{\mathrm{HS}}(\left|\boldsymbol{r}_{i,\alpha}-\boldsymbol{r}_{j,\beta}\right|)
  \label{eq:rep.energy}
\end{equation}
with the hard-sphere (HS) interaction,
\begin{equation}
  \beta u^{\mathrm{HS}}(\left|\boldsymbol{r}_{i,\alpha}-\boldsymbol{r}_{j,\beta}\right|) = 
  \begin{cases}
    \infty & \mathrm{if} \left|\boldsymbol{r}_{i,\alpha}-\boldsymbol{r}_{j,\beta}\right| < \sigma_{\alpha,\beta} \\
    0 & \mathrm{otherwise}
  \end{cases}
  \label{eq:hs.energy}
\end{equation}
where $\sigma_{\alpha,\beta}=(\sigma_{\alpha}+\sigma_{\beta})/2$ with $\alpha,\beta=h,e_1,e_2$.

Finally, we set the ear-head size ratio to $q=\sigma_e/\sigma_h = 0.85$ and the angle between the directions of the ears to $\theta_{ee}=90^{\circ}$, in order to match the values of the experiments in Ref. \citen{bib:avvisati-mickey.mouse}. However, the center-to-center distance between the ears and the head, denoted hereafter as the bond length $l/\sigma_h$, is allowed to change, and it will be used as a parameter together with the interaction range. This particle and interaction model will be referred to as MMSW (Mickey Mouse Square-Well system). Note that in the limit $l/\sigma_h = 0$ our particles correspond to HSSW particles. 

We focus on the effect of varying the ear-head distance $l/\sigma_h$ of the particle and the interaction range $\lambda$ on the gas-liquid (GL) binodal line. We compute the GL binodal using the Successive Umbrella Sampling (SUS) technique\cite{bib:virnau-sus} together with the Histogram Reweighting method\cite{bib:swendsen-hist.rew.1,bib:swendsen-hist.rew.2,bib:panagiotopoulos-hist.rew.1,bib:panagiotopoulos-hist.rew.2}. Since the SUS method has been discussed elsewhere\cite{bib:virnau-sus}, here we limit ourselves to only recalling the working scheme. The quantity of interest in the GL coexistence is $P(N)$, the probability that the system will be in a state with $N$ particles at fixed volume $V$, fixed temperature $T$, and fixed chemical potential $\mu$. This probability is unimodal above the critical temperature and assumes a typical bimodal shape, displaying gas and liquid peaks, for temperatures below the critical one. Furthermore, it can be shown that the bulk chemical potential at coexistence, at a given temperature, will be such that the area of the gas peak and the area of the liquid peak are equal. In fact, this is all we need to know to determine the coexisting chemical potential via the histogram reweighting technique\cite{bib:swendsen-hist.rew.1,bib:swendsen-hist.rew.2,bib:panagiotopoulos-hist.rew.1,bib:panagiotopoulos-hist.rew.2}. 

The probability distribution function can be computed by splitting the entire $N$ range into overlapping windows of fixed size and by performing Grand Canonical Monte Carlo (GCMC) simulations in each window. During the simulation one keeps track of how many times the system has a certain particle number $N$. which lies in between the lower and the upper limit of the window. The probability function can be reconstructed via the ``stitching'' procedure, which reads\cite{bib:virnau-sus}:
\begin{equation}
  \frac{P(N)}{P(0)} = \frac{H_0(1)}{H_0(0)}\times\frac{H_1(2)}{H_1(1)}\times\dotsm\times\frac{H_k(N)}{H_k(N-1)}
  \label{eq:pn}
\end{equation}
where we have implicitly assumed a window size of 1. From a computational point of view, Eq. \ref{eq:pn} shows us that this scheme is inherently parallel since all the ratios can be estimated from independent simulations. This is very convenient from a computer cluster perspective.

Once $P(N)$ has been calculated for a fixed temperature and (irrelevant) chemical potential, the coexistence chemical potential can be obtained by reweighting the distribution until the area under the gas peak equals the area under the liquid peak. The reweighting is carried out via the following equation\cite{bib:mccabe-hist.rev,bib:vink-hist-rew}:
\begin{equation}
  \ln P(N|\beta\mu_1) = \ln P(N|\beta\mu_0) +\beta(\mu_1-\mu_0)N
  \label{eq:hrew}
\end{equation}
In such a way, one can construct the GL coexistence envelope for particles with fixed geometry and fixed interaction range. In the following, we study how the GL coexistence curve is affected by the particle shape and particle-particle interaction range and where and how the regime of phase separation changes to self-assembly.

Once we obtain the GL coexistence curve, we attempt to estimate the critical point via a least-squares fit to the equation:
\begin{equation}
  \rho_{\pm}-\rho_c=A\left|T-T_c\right|\pm\frac{1}{2}B\left|T-T_c\right|^\beta
  \label{eq:cpfit}
\end{equation}
which stems from the law of densities and the law of the rectilinear diameter\cite{bib:lago-sw.vl.equilibrium,bib:vega-hssw,bib:mccabe-hist.rev,bib:gunton-metastable.hssw}. Here, $\pm$ stands for liquid/gas and $\beta=0.325$ is the exponent of the 3D Ising universality class. Such a procedure yields only a rough estimation and we stress that an appropriate determination of the GL critical point should involve extensive use of the finite-size scaling technique\cite{bib:wilding-fss1,bib:wilding-fss2,bib:wilding-fss3,bib:wilding-fss4}.

To explore under which conditions of particle geometry, temperature and interaction range the interparticle attractions are more important than the repulsions, we additionally compute the second virial coefficient normalised to the one of a system of hard spheres with diameter $\sigma_h$, $B^*_2=B_2/B^{\mathrm{HS}}_2$. If attractions dominate, the value of the second virial coefficient will be negative, while positive otherwise. The temperature at which the second virial coefficient vanishes, so called Boyle temperature $k_{\mathrm{B}}T_{\mathrm{Boyle}}/\varepsilon$, marks the crossing from one behaviour to another. The definition of the second virial coefficient involves the Mayer function, which takes into account the interaction between two MMSW particles\cite{bib:kubo-stat.mech,bib:hansen-simple.liquids}
\begin{equation}
  f_{ij} =\exp\left[-\beta u_{ij}\right]-1  
  \label{eq:mayer}
\end{equation}
The integral of the Mayer function over all possible positions and orientations of the two MM particles yields the second virial coefficient\cite{bib:kubo-stat.mech,bib:hansen-simple.liquids}
\begin{equation}
  B_2(k_{\textrm{B}}T/\varepsilon) = -\frac{1}{2V}\int d\boldsymbol{q}_1\,d\boldsymbol{q}_2f_{12}
    \label{eq:b2.def}
\end{equation}
where $d\boldsymbol{q}_i=d\boldsymbol{r}_i\,d\boldsymbol{\Omega}_i$ represents integration over the particle's positional and orientational degrees of freedom. Eq. \ref{eq:b2.def} can be computed via Monte Carlo integration. We place the MMSW ``1'' with fixed orientation in the center of a box with volume $V=(5\sigma_h)^3$ at a given temperature $k_{\textrm{B}}T/\varepsilon$. We then generate a number $N_c\sim\mathcal{O}(10^8)$ of random positions and orientations for the MMSW ``2'' in the same volume, and for each configuration $k$ we compute the value of the Mayer function between particle ``1'' and ``2'', $f^k_{12}$. Then, the second virial coefficient can be estimated as\cite{bib:yethiraj-sw.diatomics,bib:munao-hjdumbbells.gl}
\begin{equation}
  B_2(k_{\textrm{B}}T/\varepsilon) = -\frac{V}{2}\frac{1}{N_c}\sum_{k=0}^{N_c}f_{12}^k = -\frac{V}{2}\left\langle f_{12}\right\rangle
  \label{eq:b2.com}
\end{equation}
where $\left\langle f_{12}\right\rangle$ represents the average of Mayer function over all the configurations.
We repeat this computation for different temperatures, particle shapes and interparticle range to locate regions where attractions prevail with respect to repulsions.

\section{Results\label{sec:res}}
We have evaluated the GL coexistence curve for the MMSW model by varying the ear-head bond length $l/\sigma_h$ from $0.1$ to $0.57$, corresponding to the experimental MM particles\cite{bib:avvisati-mickey.mouse}, and the particle-particle interaction range $\lambda$ from $1.5$ to $1.02$. Following this path, we can connect the experimental MM system, for which the self-assembly behaviour has been investigated in Ref. \citen{bib:avvisati-mickey.mouse}, all the way to the HSSW limit, for which the GL coexistence has been intensively studied\cite{bib:vega-hssw,bib:lago-sw.vl.equilibrium,bib:gunton-metastable.hssw,bib:smith-hssw}. 

\subsection{\label{ssec:gl.shift}Shift of the binodal line with the particle shape}
\begin{figure*}[htb]
  \centering
  \includegraphics[scale=1.05]{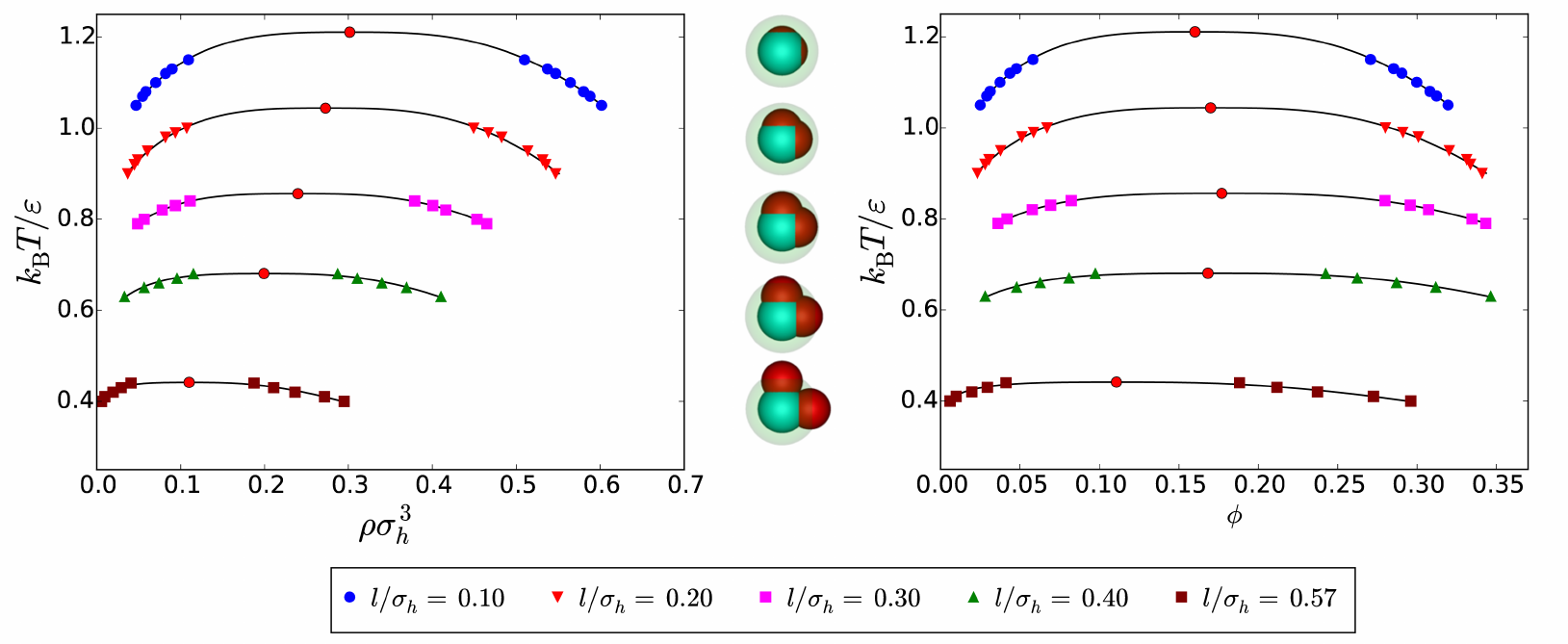}
  \caption{{\small Gas-Liquid coexistence curves for MMSW particles. We provide both the temperature $k_{\textrm{B}}T/\varepsilon$--density $\rho\sigma_h^3$, and temperature $k_{\textrm{B}}T/\varepsilon$--packing fraction $\phi$ representation. The packing fraction is calculated as $\phi=Nv_p/V$, with $v_p$ the particle volume as discussed in Appendix \ref{sec:mm.vol}. The labels stand for different increasing values of the bond length $l$. The full red dots indicate the location of the critical points by Eq. \ref{eq:cpfit}.}}
  \label{fig:mm.gl-coex}
\end{figure*}

With a fixed interaction range $\lambda=1.5$, we change the particle shape by progressively increasing the bond length distance $l/\sigma_h$ between the ears and the head of the MMSW particle. The highest value of bond length investigated in this work is set by the geometry of the experimental MM particle we want to match and reads $l/\sigma_h=0.57$. We observe that the change in the MMSW particle shape affects the position of the GL coexistence curves, as seen in Fig. \ref{fig:mm.gl-coex}. 

\begin{figure}[htb]
  \centering
  \includegraphics[scale=0.40]{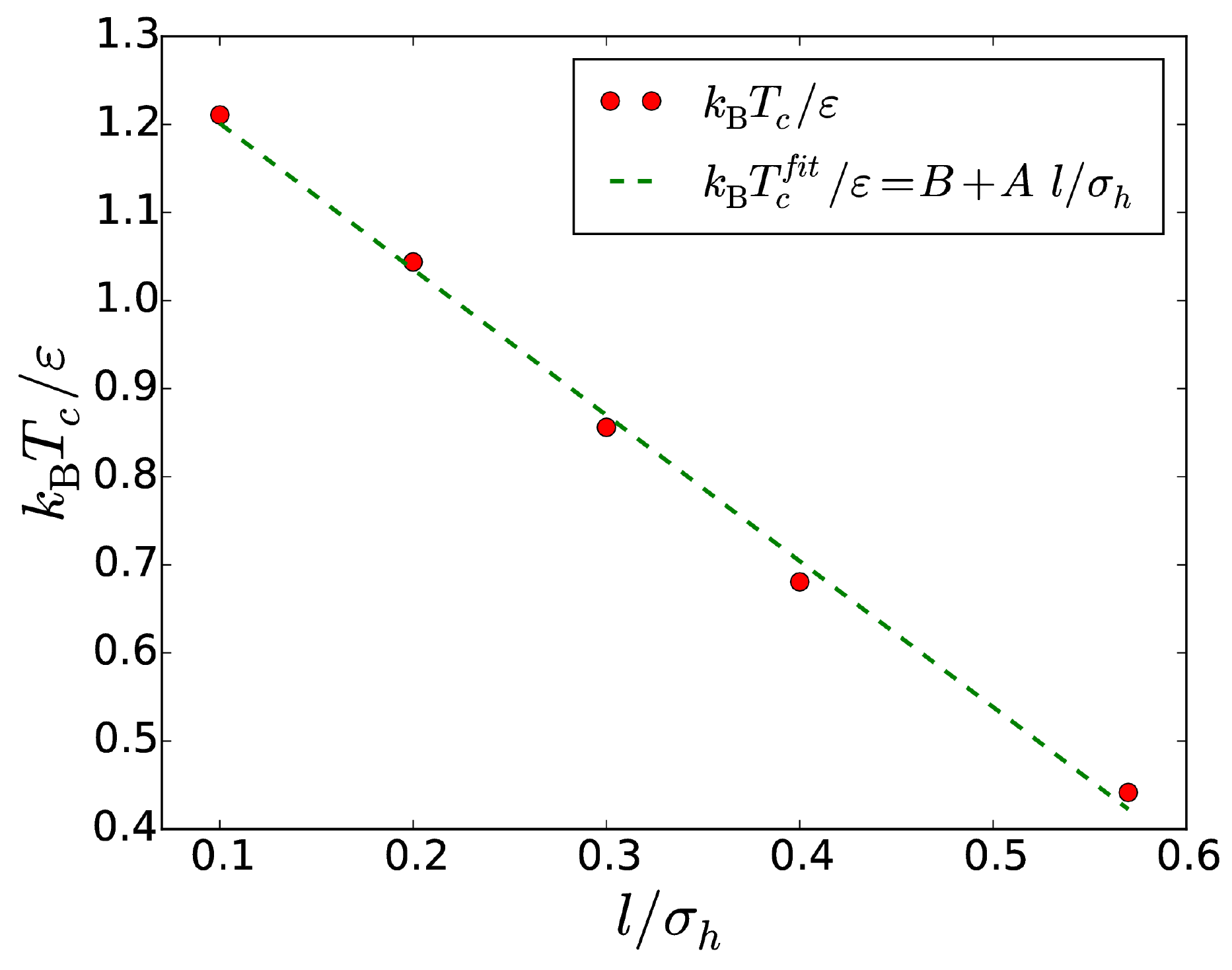}
  \caption{{\small Critical temperature $k_{\textrm{B}}T_c/\varepsilon$ (full red dots) as a function of bond length $l/\sigma_h$. The green dashed line shows the linear fit to the data for the critical temperature $k_{\textrm{B}}T_c^{fit}/\varepsilon=B+A\,l/\sigma_h$, with $B = 1.37$ and $A=-1.66$, indicating a linear relation between the bond length ``knob'' and the system's critical temperature.}}
  \label{fig:mm.tc-l}
\end{figure}

\begin{figure}[htb]
  \centering
  \includegraphics[scale=0.41]{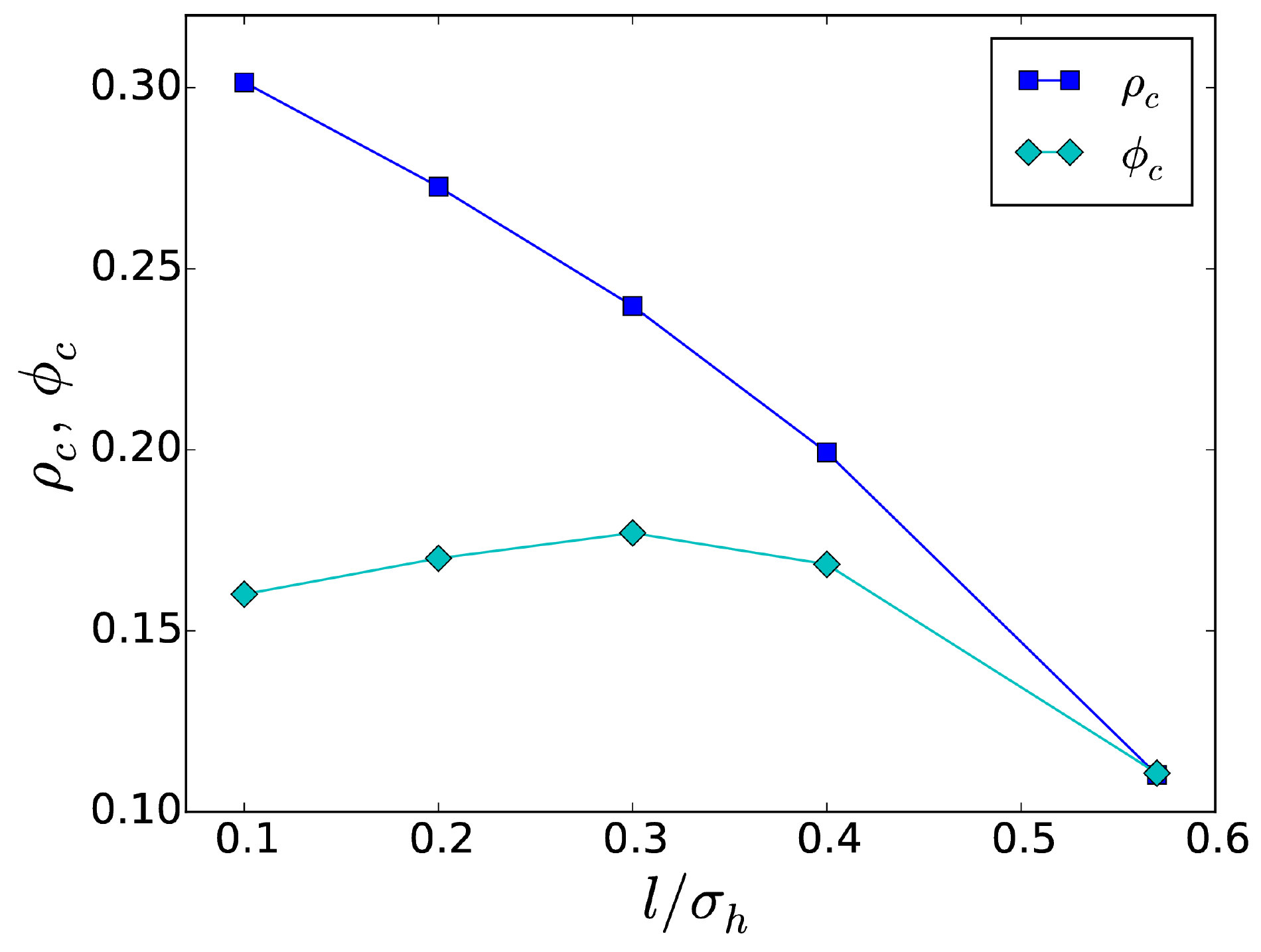}
  \caption{{\small Critical density $\rho_c$ (full blue squares) and critical packing fraction $\phi_c$ (full cyan diamonds) as a function of bond length $l/\sigma_h$. The solid curves are guides to the eye.}}
  \label{fig:mm.rc-l}
\end{figure}

We quantify the change in the binodal curves, by computing the critical parameters as function of the bond length $l/\sigma_h$ by means of a least square fit to Eq. \ref{eq:cpfit}. Going from the HSSW to the MMSW, the critical temperature $k_{\textrm{B}}T_c/\varepsilon$ shows a linear decrease upon increasing the bond length $l/\sigma_h$, as can be seen from Fig. \ref{fig:mm.tc-l}. In Fig. \ref{fig:mm.rc-l}, the critical critical density $\rho_c$ is also seen to decrease monotonically with the bond length, while for the critical packing fraction $\phi_c$ we observe a non-monotonic behaviour.

Shifts of the critical temperature to lower values have been reported in different contexts. In fact, it is well known that attractive colloidal particles confined between two parallel plates display a shift of the critical point to lower temperatures because the confinement reduces the cohesive energy of the system\cite{bib:evans-slab.criticality,bib:dijkstra-confined.coll.1,bib:dijkstra-confined.coll.2,bib:dijkstra-confined.coll.3,bib:binder-review.confinement}.
More recently, the concept of ``limited valence'' has been introduced in the context of patchy colloids\cite{bib:frenkel-kf.potential,bib:bianchi-red.valence.1,bib:bianchi-red.valence.2,bib:sciortino-lim.valence.1,bib:sciortino-lim.valence.2}. The idea is that the valence of a particle (defined as the maximum possible number of bonded nearest neighbours) affects the location of a system's critical point. Number of attractive patches\cite{bib:frenkel-kf.potential,bib:bianchi-red.valence.1,bib:bianchi-red.valence.2}, patch surface coverage\cite{bib:sciortino-lim.valence.1,bib:sciortino-lim.valence.2}, patches with tunable attractions\cite{bib:munao-patchy.dumbbells,bib:munao-janus.dumbbells}, are all examples of controlling a particle's valence. Our model suggests that changing the distance between the attractive and repulsive spheres is another way of achieving such a control. In other words, we can think of the bond length as a ``knob'' for limiting the valence of the MMSW particle. Interestingly, similar trends for the critical temperature and critical density were recently reported for a system of attractive dimers interpolating between SW-SW and HS-SW in Ref. \citen{bib:munao-janus.dumbbells}.

\begin{figure}[htb]
  \centering
  \includegraphics[scale=0.40]{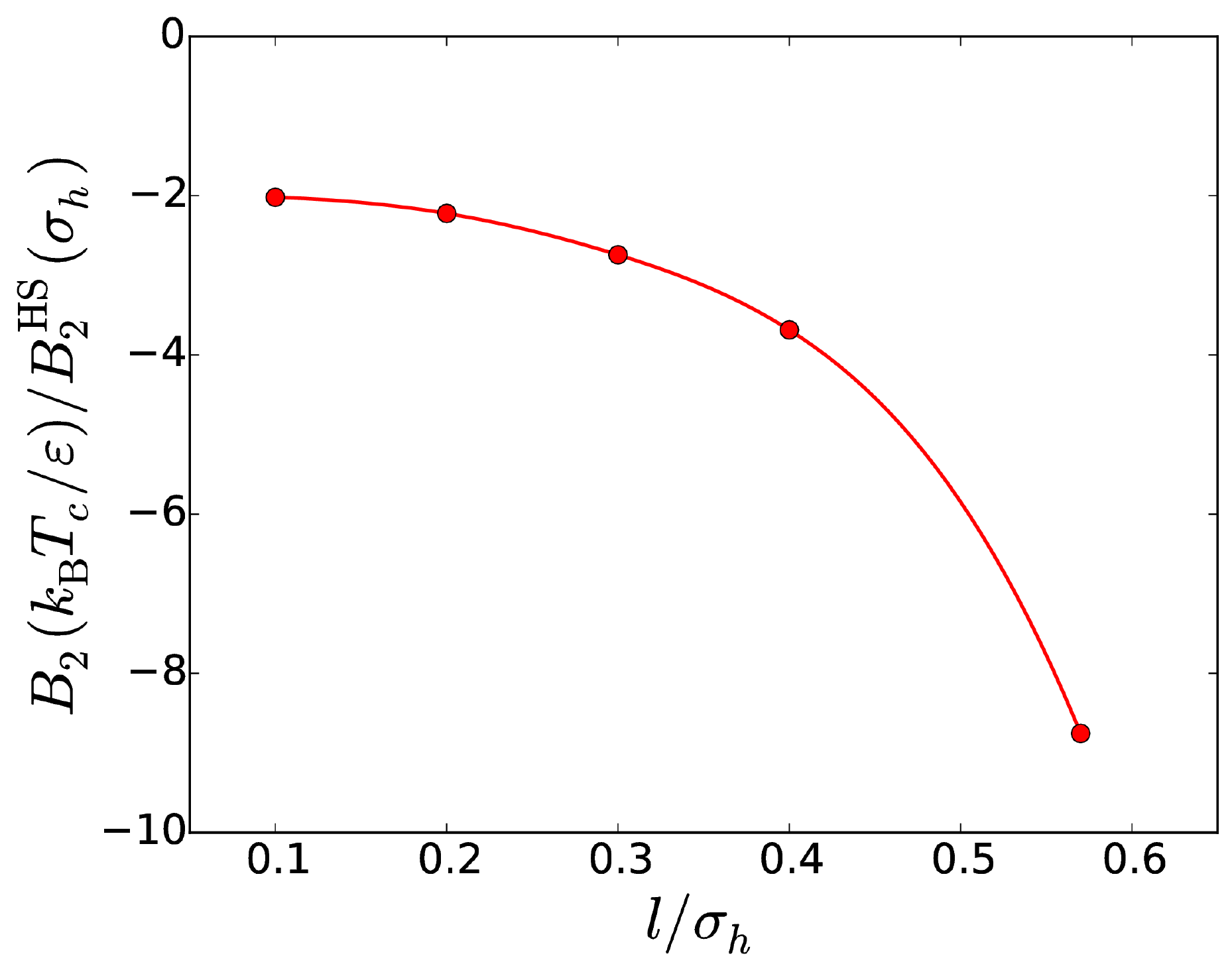}
  \caption{{\small The reduced second virial coefficient at the GL critical point as a function of the ear-head bond length $l/\sigma_h$. The red line is a guide to the eye.}}
  \label{fig:b2cr}
\end{figure}

The second virial coefficient at the critical point has been suggested as possible measure of a particle's valence in Refs. [\citen{bib:sciortino-lim.valence.1,bib:foffi-lim.valence}]. By using Eq. \ref{eq:b2.com}, we calculate the second virial coefficient $B_2$ at the critical point $(k_{\textrm{B}}T_c/\varepsilon$ for each value of the bond length $l/\sigma_h$. Our results are shown in Fig. \ref{fig:b2cr}, where we have normalised the values by the second virial coefficient of a hard sphere of diameter $\sigma_h$.  Upon increasing the bond length, the value of $B_2$ is seen to decrease (become more negative), which suggests that the limited valence picture could also hold in the system at hand.

\begin{figure}[htb]
  \centering
  \includegraphics[scale=0.43]{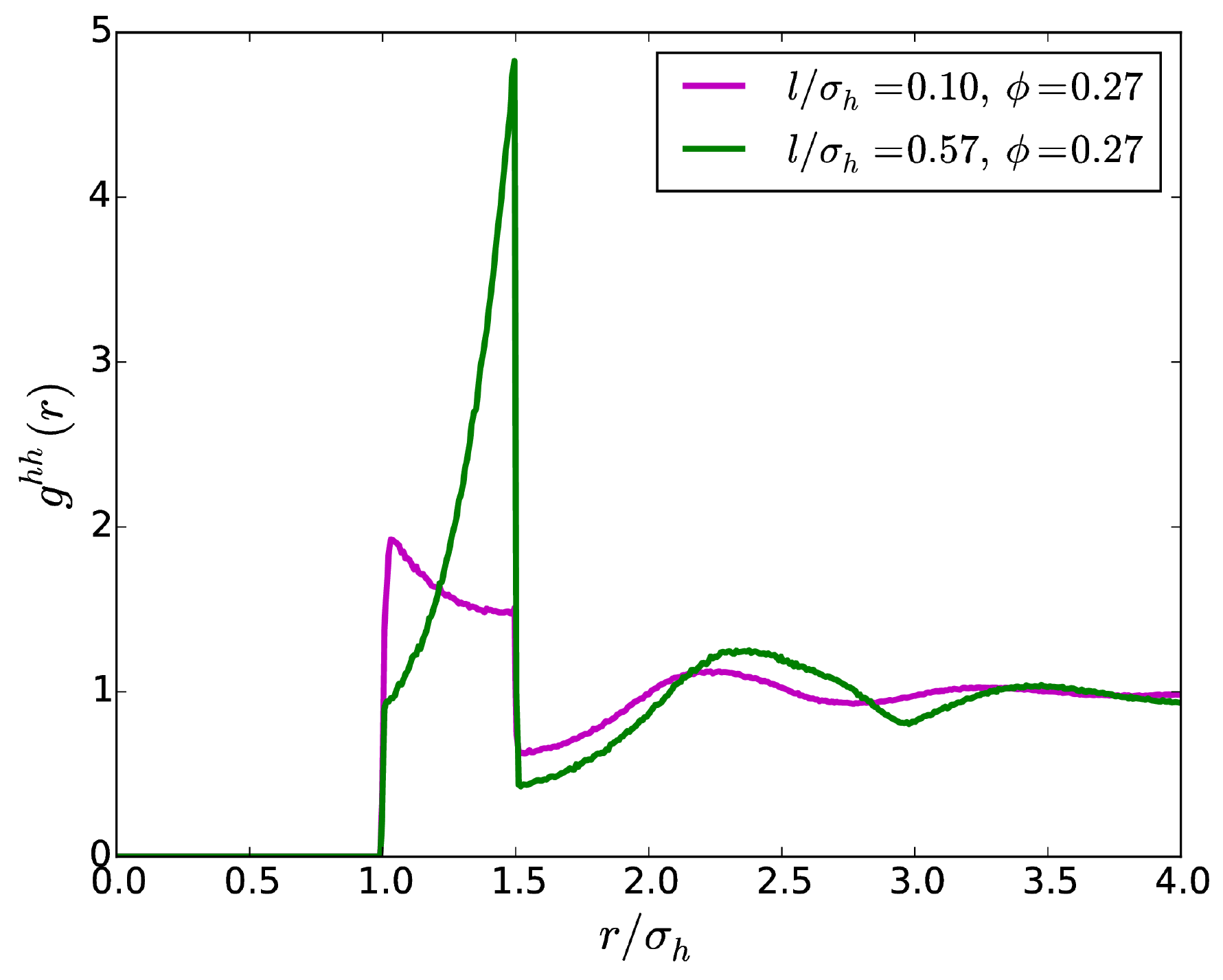}
  \caption{{\small Bulk head-head radial distribution functions $g^{hh}(r)$of the coexisting MMSW liquid at packing fraction $\phi = 0.27$ for bond length $l/\sigma_h=0.1$ (magenta curve) and $l/\sigma_h=0.57$ (green curve), as function of the distance $r$. The repulsive beads increase on average the spacing between the attractive beads.}}
  \label{fig:mm.gr.liq}
\end{figure}

It is interesting to investigate the structure of the MMSW coexisting liquid phase as function of bond length, by keeping the volume fraction fixed at $\phi = 0.27$. We do this by calculating the radial distribution function for all the MM attractive beads (heads) $g^{\mathrm{hh}}(r)$, the results being shown in Fig. \ref{fig:mm.gr.liq}, for bond length $l/\sigma_h=0.1$ and $l/\sigma_h=0.57$, respectively. We wish to remark here that we ignore the anisotropy of the particle and average over all possible particle orientations. Although both curves display features of the HSSW radial distribution functions, such as the presence of a hard-core diameter at $r/\sigma_h=1$, and a discontinuity at $r/\sigma_h=\lambda=1.5$, we see that the peak positions for the case $l/\sigma_h=0.57$ hardly resembles the ones for $l/\sigma_h=0.1$. In fact, especially the first peak of the radial distribution function moves to higher values of $r/\sigma_h$, specifically from $r/\sigma_h=1$ for $l/\sigma_h=0.1$ to $r/\sigma_h=1.5$ in the case of $l/\sigma_h=0.57$, and the same can be seen for the second and third peak. Clearly, even in the case of large interaction range $\lambda=1.5$, the presence of the ears of the MM particle has an influence on the fluid structure, pushing the heads further away from each other on average. 

\subsection{\label{ssec:sass}Interaction range-driven transition to Self-Assembly}
\begin{figure*}[htb]
  \centering
  \includegraphics[scale=0.65]{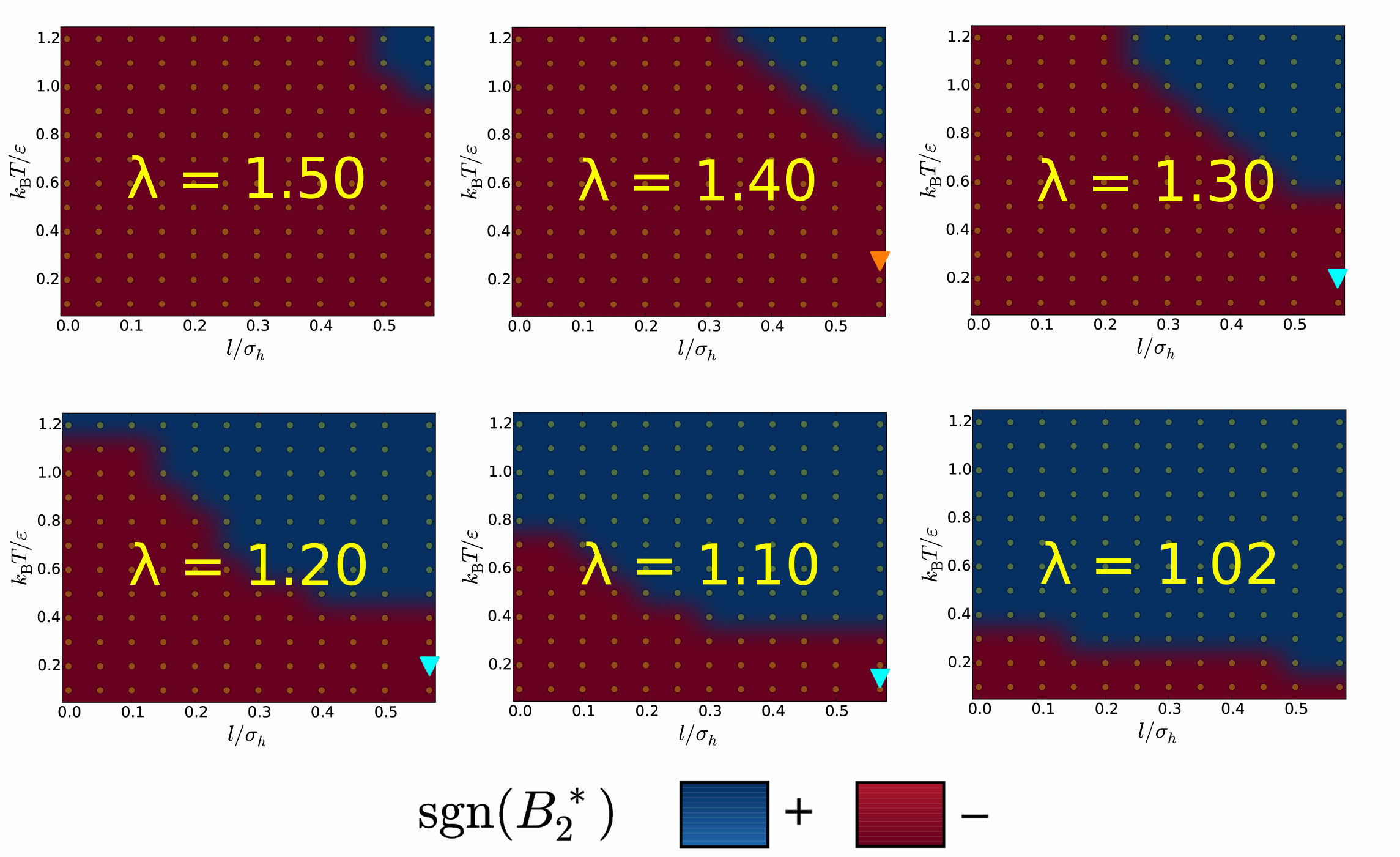}
  \caption{{\small The sign of the reduced second virial coefficient $B^*_2$ plotted in the reduced temperature $k_{\textrm{B}}T/\varepsilon$ versus bond length $l/\sigma_h$ representation, for various interaction ranges $\lambda=1.50,1.40,1.30,1.20,1.10,1.02$ as labelled. The orange triangle represent the state point where we have combined SUS and MC-NVT simulations as shown in Fig. \ref{fig:pn.selfa} and \ref{fig:range.order}, the cyan triangles represent the state points we have investigated only via MC-NVT simulations, see Fig. \ref{fig:range.order}.}}
  \label{fig:b2vl}
\end{figure*}
After having analysed how the particle shape affects the GL phase separation, we now ask ourselves how this behaviour ultimately transforms into self-assembly, which is the behaviour experimentally observed for the MM particles. In fact, the interaction range of the experimental MM particles system, i.e. $\lambda\sim 1.02$, is much smaller than the one so far considered ($\lambda\sim 1.50$). To this end, we perform a more extensive study of the second virial coefficient, where we map $B_2$ for several values of the interaction range, down to the experimentally relevant range, in the whole temperature $k_{\textrm{B}}T/\varepsilon$ and bond length plane $l/\sigma_h$. This gives us an overview of the conditions under which the attractions overcome the repulsions, and guides us in the choice of regions of the parameter space which are interesting to simulate. 

Fig. \ref{fig:b2vl} shows our results for the calculation of $B_2$ through the entire temperature $k_{\textrm{B}}T/\varepsilon$ bond length $l/\sigma_h$ plane for different values of the interaction range $\lambda = 1.50,1.40,1.30,1.20,1.10,1.02$. The results in Fig. \ref{fig:b2vl} concern the sign of the second virial coefficient $B_2$, red areas show a negative $B_2$, where attractions overcome repulsions, and blue areas refer to a positive $B_2$, where the repulsions are more important than the attractions. We remind the reader that the Boyle temperature $k_{\mathrm{B}}T_{\mathrm{Boyle}}/\varepsilon$ is defined as the temperature where the second virial coefficient $B_2$ vanishes. We also note that the case of MM geometry, which matches the experimental geometry in Ref. \citen{bib:avvisati-mickey.mouse}, is found at the right-most end of each diagram, where $l/\sigma_h=0.57$.

For $l/\sigma_h = 0.57$, the region where $B_2<0$ is seen to dramatically shrink as the interaction range is decreased. This also means that the Boyle temperature is considerably dropping, going from $k_{\mathrm{B}}T_{\mathrm{Boyle}}/\varepsilon(\lambda=1.50)\sim 0.9$ to $k_{\mathrm{B}}T_{\mathrm{Boyle}}/\varepsilon(\lambda=1.02)\sim 0.15$. It is also known that temperatures below $\sim 0.1$ can get the system stuck in kinetic traps and our calculations show that upon decreasing the interaction range $\lambda$ up to $\lambda=1.02$ the Boyle temperature gets closer to this limit. In such conditions, the temperature window where GL phase separation can be found is very small, if present at all. A confirmation of our calculation is given in Ref. \citen{bib:avvisati-mickey.mouse}, where it is mentioned that the temperature window between clustering and kinetic trapping is very narrow.

\begin{figure}[htb]
  \centering
  \includegraphics[scale=0.41]{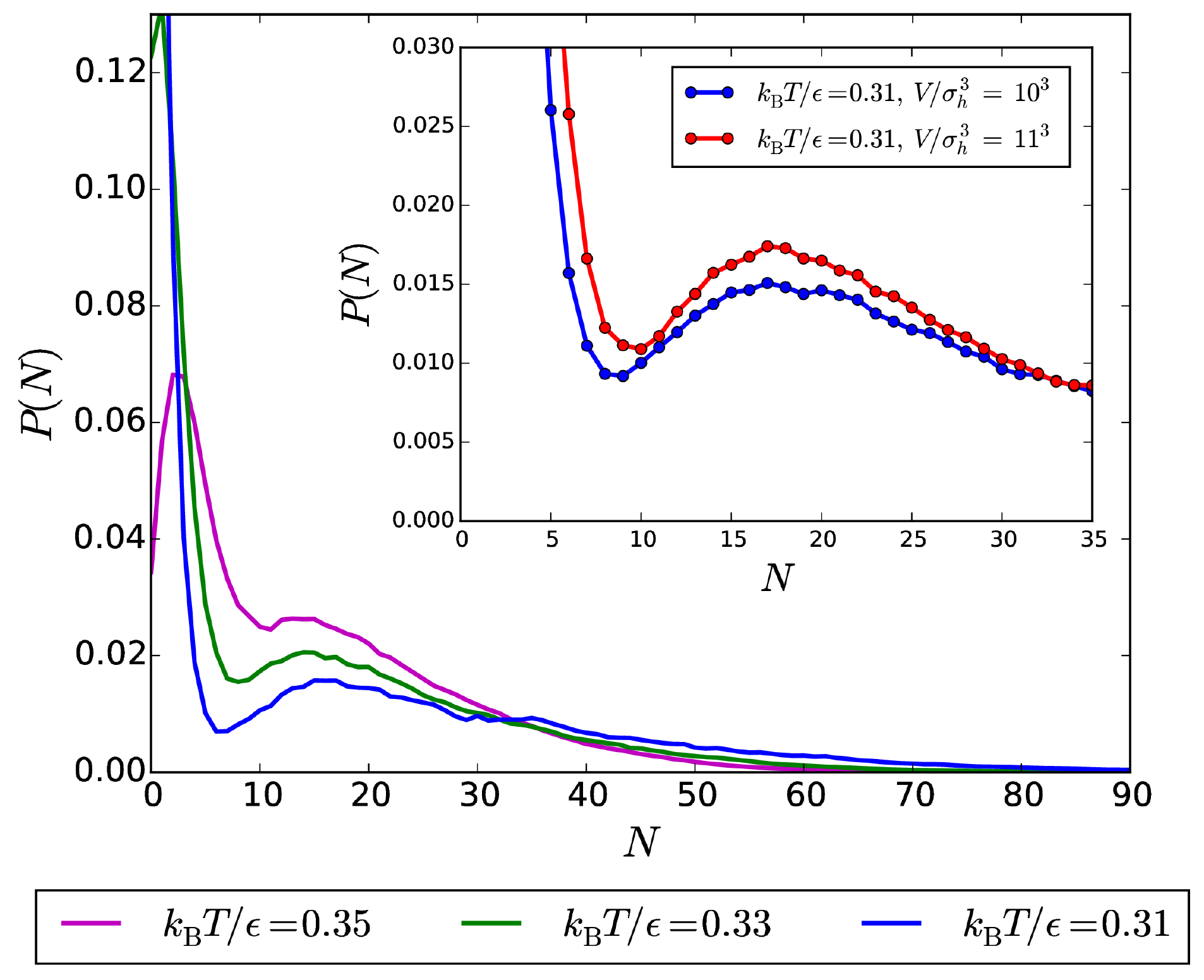}
  \caption{{\small Probability of finding $N$ particles in the system for bond length $l/\sigma_h=0.57$, interaction range $\lambda = 1.40$ and temperature as labelled. All the distributions are reweighted as to highlight the peaked region present for particle number $N \in [10,40]$. Inset: finite size study performed at temperature $k_{\mathrm{B}}T/\varepsilon=0.31$ confirming that the second peak is always located in the same window of particle number $N$.}}
  \label{fig:pn.selfa}
\end{figure}

Based on the second virial coefficient analysis, the largest temperature window where it is in principle still possible to locate a GL phase separation is $\lambda = 1.40$, besides the already investigated case of $\lambda = 1.50$. We have therefore calculated $P(N)$ for a temperature range of $k_{\mathrm{B}}T/\varepsilon \in [0.31, 0.35]$, the resulting distributions being shown in Fig. \ref{fig:pn.selfa}. 

For all investigated temperatures, the distributions appear to develop a shoulder at higher particle number $N$ than the the first peak corresponding to the gas phase. Upon increasing the system volume, this second peak is shown to remain at the same location, as shown in the inset of Fig. \ref{fig:pn.selfa}, and is therefore associated with clusters developing in the system. This suggests that a self-assembly process preempts the macroscopic GL phase separation, either by shifting the transition to much lower temperatures than the investigated ones, or by destabilising the liquid with respect to the micellar fluid, as pointed out in previous works\cite{bib:camp-gl.dipolar,bib:rovigatti-gl.dipolar.sph,bib:dussi-gl.dumbbells,bib:munao-janus.dumbbells}. Thus, even for the best case scenario of $\lambda=1.40$, we have not found evidence of a GL phase separation, but instead we find indication of spontaneously formed aggregates in the system.

\begin{figure}[htb]
  \centering
  \includegraphics[scale=1.6]{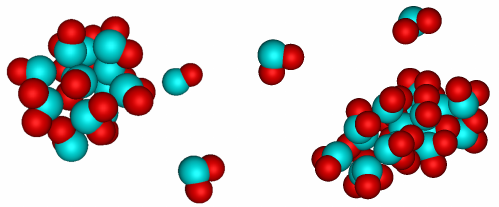}
  \caption{{\small Disordered clusters of MMSW particles for bond length $l/\sigma_h=0.57$, interaction range $\lambda = 1.40$ and temperature $k_{\mathrm{B}}T/\varepsilon=0.31$. (left) $N=15,\,\phi=0.011$, (right) $N=20,\,\phi=0.015$.}}
  \label{fig:dis.micelles}
\end{figure}

The snapshots of the system for particle numbers within the micellar peak of $P(N)$ ($N=15$, $N=20$, $N=20$), shown in Fig. \ref{fig:dis.micelles}, reveal the disordered microscopic configurations of the clusters associated with the relatively high interaction range $\lambda = 1.40$.

\begin{figure}[htb]
  \includegraphics[scale=1.35]{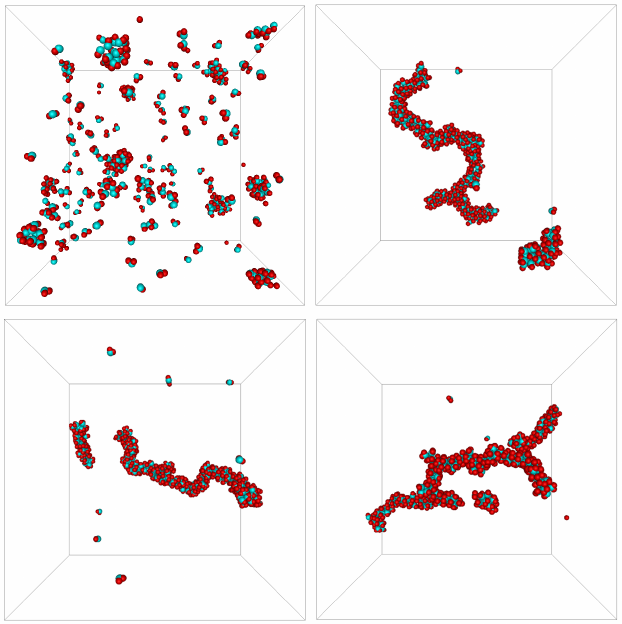}
  \caption{{\small Snapshots from MC NVT simulations at fixed packing fraction $\phi=Nv_p/V = 0.006$ and bond length $l/\sigma_h=0.57$ for different interaction ranges $\lambda$ and temperatures $k_{\textrm{B}}T/\varepsilon$ as described by the colored triangles in Fig. \ref{fig:b2vl}. Top left: amorphous clusters for $\lambda = 1.40$ and $k_{\textrm{B}}T/\varepsilon = 0.310$, in agreement with the $P(N)$ calculated at the same temperature in Fig. \ref{fig:pn.selfa}; Top right: branched tube-like structures appearing at $\lambda = 1.30$ and $k_{\textrm{B}}T/\varepsilon = 0.200$; Bottom left: self-assembled tubes at $\lambda = 1.20$ and $k_{\textrm{B}}T/\varepsilon = 0.200$; Bottom right: branched tubes at $\lambda = 1.10$ and $k_{\textrm{B}}T/\varepsilon = 0.143$. For all the panels, $N=256$ and $V=35\sigma_h^3$}.}
  \label{fig:range.order}
\end{figure}

For lower values of the interaction range $\lambda$, we select few state points, as indicated in Fig. \ref{fig:b2vl}, and investigate the system via MC simulations in the NVT ensemble, due to computational reasons. The final configurations of the NVT simulations for $\lambda = 1.40, 1.30, 1.20, 1.10$ are reported in Fig. \ref{fig:range.order}. The panel for a long ranged interaction $\lambda=1.40$ (top left) displays a micellar fluid with clusters of size comparable to the peak in the computed $P(N)$. On the other hand, the remaining panels for shorter-ranged interactions show self-assembled, tube-like structures. Once the self-assembly regime has been reached, starting at $\lambda=1.40$, the shape of the resulting structures can be tuned by decreasing the interaction range. The backbone of the tubes becomes more close packed upon decreasing the interaction range, and the branching is suppressed for $\lambda\leq 1.05$. Thus, structures as the Bernal spirals, which were reported in Ref. \citen{bib:avvisati-mickey.mouse}, are eventually recovered upon shrinking the interaction range to values lower than $\lambda = 1.05$.

The above findings highlight the role of the range in the self-assembly of MM particles: shrinking the interaction range first destabilises the GL phase separation with respect to self-assembly, and then favours the formation of increasingly ordered structures, eventually recovering those found in Ref. \citen{bib:avvisati-mickey.mouse}.

In a very recent work, the fluid phase behaviour of trimers with one soft, attractive bead and two soft, repulsive beads has been investigated\cite{bib:hatch-trimer.selfa}. There are substantial differences with our investigation, which we summarise in the following. Firstly, the particle model employed here uses hard-core particles, whereas the model in Ref. \citen{bib:hatch-trimer.selfa} makes use of soft-core particles. Secondly, the shape of their trimer is slightly different from our MM particle, since we keep the angle between the repulsive beads fixed. Thirdly, their methodology is fundamentally different, although it aims to compute the same quantity $P(N)$. Lastly, due to our SW potential, we can explicitly control the interaction range between two particles, and track down the evolution of the self-assembly with it. Notwithstanding these differences, it is interesting to compare our results to Ref. \citen{bib:hatch-trimer.selfa}. Concerning the shift of the binodal line with the bond length distance, both works have found a linear scaling. The slopes of the scaling are comparable, our estimate being $a=-1.66$, see the fit in Fig. \ref{fig:mm.tc-l}, against the reported $a=-1.40$. However, the actual values of the critical points for specific bond lengths do not match with each other, the discrepancy being ascribed to the differences in the particle geometry and particle-particle interaction. Furthermore, by very different analyses, both works agree on the region where the tube-like structures are expected to occur for the experiments reported in Ref. \citen{bib:avvisati-mickey.mouse} -- roughly for temperatures $k_{\textrm{B}}T/\varepsilon < 0.15$.

\section{\label{sec:summary}Conclusions}
In this work we have investigated, via Monte Carlo simulations, how the combined actions of particle geometry and particle-particle interaction range affect the gas liquid phase separation. We have also addressed the role of the interaction range onto the self-assembly process. 

Starting from the well-known HSSW model, we have mapped out the different binodal curves for a family of Mickey Mouse particles with increasing steric repulsion. We have found that there is a linear relation between the critical temperature $k_{\textrm{B}}T_c/\varepsilon$ of the binodal curves and the bond length $l/\sigma_h$, which controls the influence of the repulsive spheres on the overall interaction. We interpreted this phenomenon in the ``limited valence'' paradigm and documented it by calculating the value of the second virial coefficient at the critical temperature $B^*_2(k_{\textrm{B}}T_c/\varepsilon)$ as function of bond length $l/\sigma_h$. This is a possible measure of the effective valence of the system and indeed we have proven that this value decreases as the valence is reduced, which is controlled in this case via the bond length.

The transition from macroscopic GL phase separation and self-assembly is driven by the interaction range, which has also an important effect on the Boyle temperature $T_{\mathrm{Boyle}}$ of the system. In fact, as the interaction range is reduced, both the Boyle temperature decreases and the phase separation process is preempted by a self-assembly process. Microscopic configurations from the SUS simulations, complemented with snapshots from NVT simulations, have shown that the structure of the self assembled clusters changes from disordered to more ordered, branched tube-like structures. Upon further decrease in the range, these structures are seen to transform into the very ordered ones as reported in Ref. \citen{bib:avvisati-mickey.mouse}.

Our results highlight the role of particle geometry and interparticle interaction range in the competition between phase separation and self-assembly for Mickey Mouse particles and, in general, trimers with one attractive sphere and two repulsive spheres. This is of direct interest for experiments where trimers are used as basic building blocks, where -- as we have documented -- both the trimer exact shape and the trimer-trimer interaction range are expected to play a significant role. At a more general level, we have shown how tuning a particle's valence can be achieved by controlling the bond length. This conclusion is expected to hold also in other particle geometries as long as the particles are composed of different beads, e.g. colloidal dimers with one rough (repulsive) and one smooth (attractive) beads.

\section*{Acknowledgements}
This work is part of the research programme of the Foundation for Fundamental Research on Matter (FOM), which is part of the Netherlands Organisation for Scientific Research (NWO). M.D. acknowledges financial support from a NWO-VICI grant. G.A. thanks E. Luijten and F. Sciortino for fruitful discussions. The authors thank J.R. Edison and Z. Preisler for critically reading the manuscript. 

\appendix
\section{\label{sec:mm.vol}MM particle volume calculation}
The volume of a MM particle $v_p$ can be calculated exactly via MC integration, and in an approximated fashion considering the amount of overlap volume between the two ears and the head.  In fact, from geometrical considerations, it holds that
\begin{equation}
  v_{p} = V_h + 2V_e - 2V_{h,e} - V_{e,e}
  \label{eq:mm.vt}
\end{equation}
where $V_h$ is the volume of the attractive sphere, $V_e$ the volume of the repulsive sphere, $V_{h,e}$ is the head-ear overlap volume, and $V_{e,e}$ is the ear-ear overlap volume.
Since, as first approximation, the two ears in a MM particle do not overlap with each other in the region outside the head, their overlap volume $V_{e,e}$ is small compared to the head-ear one $V_{h,e}$ so the latter is the only quantity we need to calculate, which is given by 
\begin{align}
  V_{h,e} & \equiv V(R_h,R_e,l) = \nonumber\\
  & = \frac{\pi\left(R_h+R_e-l\right)^2}{12l} \times \nonumber\\
  & \times \left(l^2+2lR_e-3R_e^2+2lR_h+6R_eR_h-3R_h^2\right)
  \label{eq:mm.vhe}
\end{align}
where $R_\alpha=0.5\sigma_\alpha,\;\alpha=h,e_1,e_2$ are the radii of the attractive and the repulsive spheres and $l$ is the center to center distance of the spheres.

\begin{figure}[htb]
  \centering
  \includegraphics[scale=0.42]{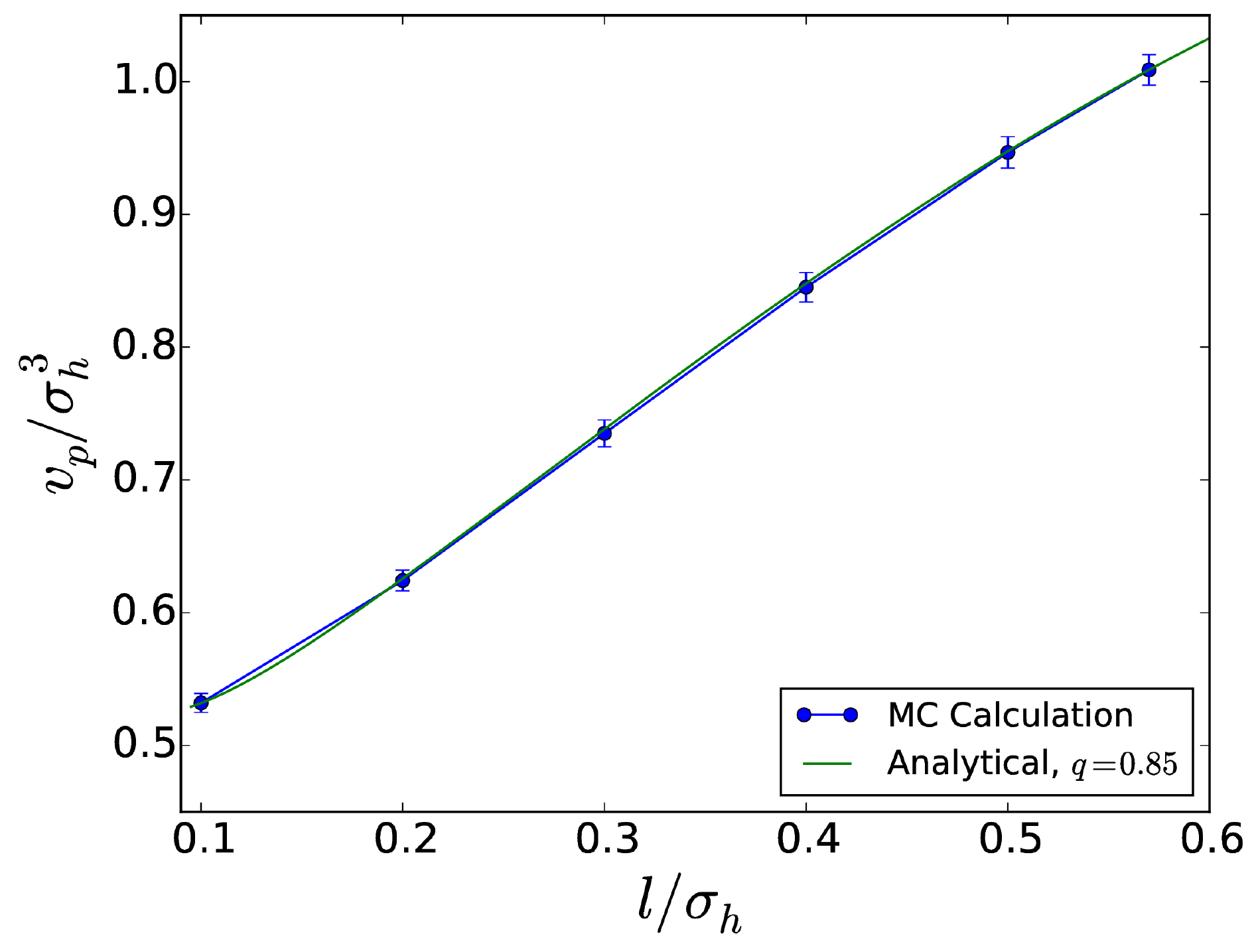}
  \caption{{\small Volume of a MMSW particle as a function of bond length $l/\sigma_h$ for size ratio $q=0.85$. The solid green line is the prediction from Eq. \ref{eq:mm.vt} with $q=0.85$, and the blue dots are the results from the MC calculations according to Eq. \ref{eq:mm.vt.mc}}. The errors are multiplied by 10 for visualisation purposes.}
  \label{fig:vt.comp}
\end{figure}

The Monte Carlo integration is similar to a MC calculation of $\pi$, and we discuss it in the following. We place a MM particle in a box with fixed volume $V$, big enough to enclose the particle. We then generate a random position in the box and check whether it overlaps with the MM particle. If the number of generated random positions is $N_{trial}\sim\mathcal{O}(10^8)$ and the number of hits on the MM particles is $N_{hit}$, then the MM particle volume is given by
\begin{equation}
  v_p \simeq V\frac{N_{hit}}{N_{trial}}
  \label{eq:mm.vt.mc}
\end{equation}
The results of the calculations for the volume of a MMSW particle are shown in Fig. \ref{fig:vt.comp}, for the ratio studied in this paper $q=0.85$, as a function of bond length $l/\sigma_h$. The MC integration and the theoretical calculation are almost indistinguishable.

\bibliography{paper-mmgl} 

\end{document}